\documentclass[10pt,conference]{IEEEtran}
\usepackage[T1]{fontenc}
\usepackage[noadjust]{cite}
\usepackage{amsmath}
\usepackage{amssymb}
\usepackage{amsthm}
\usepackage{mathtools}
\usepackage{dsfont}
\usepackage{algorithm}
\usepackage[noend]{algpseudocode}
\usepackage{bm}
\usepackage{multirow}
\usepackage{comment}
\usepackage{booktabs}
\usepackage[caption=false,font=footnotesize]{subfig}

\DeclareMathOperator*{\argmax}{arg\,max}

\DeclareMathOperator{\E}{\mathbb{E}}
\usepackage{pgfplots}
\pgfplotsset{compat=newest,
legend image code/.code={
\draw[mark repeat=2,mark phase=2]
plot coordinates {
(0cm,0cm)
(0.15cm,0cm)        
(0.3cm,0cm)         
};%
}}

\usepgfplotslibrary{fillbetween}
\usetikzlibrary{patterns,arrows,plotmarks,matrix}
\usepgfplotslibrary{groupplots}
\pgfdeclarelayer{background}
\pgfsetlayers{background,main}
\usetikzlibrary{automata,positioning}
\usetikzlibrary{decorations}
\usetikzlibrary{shapes.arrows}
\usetikzlibrary{tikzmark}
\usetikzlibrary{calc}
\usetikzlibrary{decorations.markings}
\usetikzlibrary{shapes.geometric}
\algrenewcommand\algorithmicindent{10pt}
\usepgfplotslibrary{colorbrewer}
\usepackage{glossaries}    

\newacronym[plural=MDPs,firstplural=Markov decision processes (MDPs)]{mdp}{MDP}{Markov decision process}
\newacronym{iot}{IoT}{Internet of Things}
\newacronym{fec}{FEC}{forward error correction}
\newacronym{snr}{SNR}{signal to noise ratio}
\newacronym{iid}{iid.}{independent and identically distributed}
\newacronym{csi}{CSI}{channel state information}
\newacronym{epdf}{EPDF}{empirical probability density function}
\newacronym{harq}{HARQ}{hybrid automated repeat request}
\newacronym{arq}{ARQ}{automated repeat request}
\newacronym{dharq}{D-HARQ}{dynamic HARQ}
\newacronym{ack}{ACK}{acknowledgment}
\newacronym{nack}{NACK}{negative acknowledgment}
\newacronym{dqn}{DQN}{deep Q-network}
\newacronym{ldpc}{LDPC}{low-density parity check}
\newacronym{drl}{DRL}{deep reinforcement learning}
\newacronym{rl}{RL}{reinforcement learning}
\newacronym{nbiot}{NB-IoT}{narrowband IoT}
\newacronym{awgn}{AWGN}{additive white Gaussian noise}
\newacronym{bp}{BP}{belief propagation}
\newacronym{crc}{CRC}{cyclic redundancy check}
\newacronym{qam}{QAM}{quadrature amplitude modulation}
\newacronym[plural=KPIs,firstplural=key performance indicators (KPIs)]{kpi}{KPI}{key performance indicator}
\newacronym{qpsk}{QPSK}{quadrature phase-shift keying}
\newacronym{raf}{RAF}{reinforcement-based adaptive feedback}
\newacronym{ss}{SS}{static single}
\newacronym{st}{ST}{static tapered}
\newacronym{ta}{TA}{threshold-based adaptive}
\newacronym{uder}{UDER}{undetected error rate}
\newacronym{mr}{MR}{multiplicatively repeated}
\newacronym{sdo}{SDO}{sequential differential optimization}
\newacronym{rf}{RF}{radio frequency}
\newacronym{gf}{GF}{Galois field}
\newacronym{relu}{ReLU}{rectified linear unit}
\newacronym{cdf}{CDF}{cumulative distribution function}
\newacronym{pdf}{PDF}{probability density function}
\newacronym{cc}{CC}{chase combining}
\newacronym{ir}{IR}{incremental redundancy}
\newacronym{rbharq}{RB-HARQ}{reliability-based HARQ}
\newacronym{mds}{MDS}{maximum distance separable}
\newacronym{bsc}{BSC}{binary symmetric channel}
\newacronym[plural=LLRs,firstplural=log-likelihood ratios (LLRs)]{llr}{LLR}{log-likelihood ratio}
\newacronym[plural=DNNs,firstplural=deep neural networks (DNNs)]{dnn}{DNN}{deep neural network}

\makeatletter
\newcommand{\shorteq}{%
  \settowidth{\@tempdima}{+}
  \resizebox{\@tempdima}{\height}{=}%
}
\makeatletter
\newcommand\notsotiny{\@setfontsize\notsotiny\@vipt\@viipt}
\makeatother

\errorcontextlines\maxdimen

\definecolor{color1}{HTML}{0011af}
\definecolor{color2}{HTML}{8819a0}
\definecolor{color3}{HTML}{bf418d}
\definecolor{color4}{HTML}{e37076}
\definecolor{color5}{HTML}{f9a256}
\definecolor{color6}{HTML}{FF0000}
\definecolor{color7}{HTML}{009F6B}
\definecolor{color8}{HTML}{00CC99}

\def \gfwidth {\linewidth}
\def \gfheight {0.9\linewidth}

\newcommand{\code}{\mathcal{C}}

\IEEEoverridecommandlockouts

\begin{document}
\title{Learning-Based Rich Feedback HARQ for Energy-Efficient {Uplink} Short Packet Transmission}

\author{%
\IEEEauthorblockN{Martin V. Vejling\IEEEauthorrefmark{1}, Federico Chiariotti\IEEEauthorrefmark{2}, Anders E. Kal\o{}r\IEEEauthorrefmark{3}, Deniz G\"und\"uz\IEEEauthorrefmark{4}, Gianluigi Liva\IEEEauthorrefmark{5}, Petar Popovski\IEEEauthorrefmark{1}
\IEEEauthorblockA{\IEEEauthorrefmark{1}Dept. of Electronic Systems, Aalborg University, Denmark (\{mvv,petarp\}@es.aau.dk)}
\IEEEauthorblockA{\IEEEauthorrefmark{2}Dept. of Information Engineering, University of Padova, Italy (chiariot@dei.unipd.it)}
\IEEEauthorblockA{\IEEEauthorrefmark{3}Dept. of Information and Computer Science, Keio University, Japan (aek@keio.jp)}
\IEEEauthorblockA{\IEEEauthorrefmark{4}Dept. of Electrical and Electronic Engineering, Imperial College London, UK (d.gunduz@imperial.ac.uk)}
\IEEEauthorblockA{\IEEEauthorrefmark{5}Inst. of Communications and Navigation, German Aerospace Center (DLR), Wessling, Germany (gianluigi.liva@dlr.de)}}\thanks{This work was partly funded by the Villum Investigator Grant ``WATER'' financed by the Villum Foundation, Denmark. The work of F.~Chiariotti was financed by the European Union under the Italian National Recovery and Resilience Plan of NextGenerationEU, under the ``SoE Young Researchers'' grant REDIAL (SoE0000009). The work of A.~E.~Kal\o{}r was supported by the Independent Research Fund Denmark (IRFD) under Grant 1056-00006B.}
}

\maketitle

\begin{abstract}
The trade-off between reliability, latency, and energy efficiency is a central problem in communication systems. Advanced \gls{harq} techniques reduce retransmissions required for reliable communication but incur high computational costs. Strict energy constraints apply mainly to devices, while the access point receiving their packets is usually connected to the electrical grid. Therefore, moving the computational complexity from the transmitter to the receiver may provide a way to improve this trade-off. We propose the \gls{raf} scheme, a departure from traditional single-bit feedback \gls{harq}, introducing adaptive \emph{rich feedback} where the receiver requests the coded retransmission of specific symbols. Simulation results show that \gls{raf} achieves a better trade-off between energy efficiency, reliability, and latency, compared to existing \gls{harq} solutions. Our \gls{raf} scheme can easily adapt to different modulation schemes and can also generalize to different channel statistics.
\end{abstract}

\glsresetall

\section{Introduction}
Over the past few years, \gls{iot} systems have become ubiquitous: factories, buildings, farms, and urban environments now have extensive networks of simple \gls{iot} devices, which {are used to} collect all sorts of data, improving energy efficiency and informing economic and policy decisions~\cite{zhou2020near,guo2021enabling}. Most \gls{iot} devices have {very} long design lifespans, often measured in years or decades~\cite{singh2020energy}, {and thus their main limitation is the energy consumption, since replacing batteries is an expensive operation in many cases.} As a result, energy efficiency is a significant concern in \gls{iot} research, and several low-energy protocols have been developed~\cite{mekki2019comparative}.

In addition to the energy consumption, \gls{iot} devices are typically characterized by short packet transmissions, primarily in the uplink~\cite{guo2021enabling}. Furthermore, some applications, such as industrial and environmental monitoring, require relatively high reliability. To achieve this, \gls{iot} devices typically implement retransmission schemes such as \gls{harq}~\cite{ahmed2021hybrid}, in which the receiver combines all previous transmissions to decode a message. As such, \gls{harq} dynamically adjusts the transmission rate to the instantaneous channel conditions. While extra transmissions come at the cost of increased delay and energy consumption at the device, only retransmitting information when it is necessary contributes to an average reduction in the total energy consumption.

Overall, \gls{harq} introduces a fundamental trade-off between reliability, latency, and energy consumption: (\emph{i}), a low coding rate in the first transmission leads to fewer rounds of \gls{harq}, and hence a lower latency, but requires more energy; (\emph{ii}), transmitting few symbols in each round leads to higher latency and lower reliability, but is more energy efficient, as the {energy-constrained} device transmits the exact number of symbols needed to decode the {message}~\cite{avranas2018energy}. On the other hand, multiple rounds of feedback can also lead to a high energy consumption, as the reception of the feedback may require almost as much energy as active transmission~\cite{andres2019analytical}. Computation is another aspect that cannot be neglected: complex schemes that put a significant computational load on the transmitter side can deplete batteries just as quickly as inefficient transmission~\cite{li2019green}.

In this paper, we propose a new \gls{harq} strategy, termed \gls{raf}, designed to minimize the total energy consumption of a device transmitting in the uplink while maintaining a relatively low latency and high reliability.
The main idea behind \gls{raf} is to move the complexity from the transmitter to the receiver, which does not have the same energy limitations, and to adaptively and optimally exploit \gls{rbharq}~\cite{Shea2002:Reliability} by informing the transmitter exactly which parts of the packet are the most ambiguous. This way, the transmitter can construct retransmissions that directly target these portions of the packet, thereby increasing the likelihood that the retransmission will contribute to correct decoding of the message, while keeping the complexity at the device low.

The main contributions can be summarized as follows:
\begin{enumerate}
    \item 
    We propose \gls{raf}, an \gls{rbharq} scheme, which learns the optimal mapping of the posterior probabilities, obtained from the \gls{bp} decoder, to the list of symbols for retransmission. Since the complex interaction between the encoder and decoder of such a code makes it hard to describe the feedback policy analytically, we model the feedback design problem as a \gls{mdp}, and use \gls{drl} to optimize the trade-off between the energy efficiency and latency while guaranteeing high reliability.
    \item The proposed scheme is compared to existing \gls{harq}, \gls{dharq}, and \gls{rbharq} solutions in the context of \gls{mr} non-binary \gls{ldpc} codes with different modulation schemes and channels.
\end{enumerate}
Our simulations show that \gls{raf} can outperform the existing \gls{harq}, \gls{dharq}, and \gls{rbharq} policies by improving all \glspl{kpi} simultaneously.
Specifically, with an \gls{awgn} channel, we are able to reduce the undetected error rate by a factor of $4$, while latency is decreased by $4\%$ and energy efficiency is increased by $1\%$ with respect to the optimized benchmarks. For a block fading channel, energy efficiency improves by up to $4\%$ with the same latency.

The rest of this paper is organized as follows. Sec.~\ref{sec:system_model} introduces the system model, including the energy model, and the \gls{harq} scheme considered in this work.
Sec.~\ref{sec:soft_decoding} then presents the proposed \gls{raf} scheme, describing the \gls{mdp} formulation and the adopted \gls{dqn} approach. Numerical results are presented in Sec.~\ref{sec:sim}, and Sec.~\ref{sec:conc} concludes the paper and presents some future research directions.

\section{System Model}\label{sec:system_model}
We consider an energy-constrained device transmitting a $K$-bit message $m\in\{0,1,\ldots,2^{K}-1\}$ in the uplink to a base station over a wireless channel with minimal energy usage, high reliability, and low latency. The receiver has a relaxed energy constraint and high computational capabilities. Transmission occurs in rounds $t=0,1,\ldots,T_{\mathrm{max}}-1$, involving uplink data transmission of a variable number of code symbols $L_t$ over $\hat{L}_t$ channel uses, and a fixed number of downlink feedback symbols $L_{\mathrm f}$ over $\hat{L}_{\mathrm f}$ channel uses. For instance, with $M$-ary \gls{qam} and codewords over the order-$q$ \gls{gf} $\mathbb{F}_q$, the number of uplink channel uses in the $t$-th round is $\hat{L}_t = \frac{\log_2(q)}{\log_2(M)}L_t$, and the communication rate after the $t$-th round is $R_t = K/(\sum_{i=0}^t\hat{L}_i)$ bits per channel use. We will assume that $q\le M$ so that each code symbol produces $\frac{\log_2(q)}{\log_2(M)}\ge 1$ channel symbols, but the general framework can be applied to the alternative case as well. The same modulation scheme is used for downlink, so that $\hat{L}_{\mathrm{f}}=\frac{\log_2(q)}{\log_2(M)}L_{\mathrm{f}}$.
To facilitate channel estimation, each frame (both uplink and downlink) also contains a preamble of $\hat{L}_p$ channel uses. 

Uplink symbols $\mathbf{x}_t \in \mathbb{C}^{\hat{L}_t}$ have normalized power and are transmitted over a block fading channel, yielding received signals $\mathbf{y}_t=\beta_t\mathbf{x}_t+\mathbf{w}_t$, where $\beta_t\overset{iid}{\sim} {\rm P}_{\beta}$ is the instantaneous fading coefficient \gls{iid} from some distribution ${\rm P}_{\beta}$, and $\mathbf{w}_t\sim \mathcal{C}\mathcal{N}(\mathbf{0}, \sigma^2\mathbf{I})$ is \gls{awgn}. The instantaneous \gls{snr} is given by $\text{SNR}_t = \frac{|\beta_t|^2}{\sigma^2}$. Feedback is assumed to be error-free since the receiver can transmit at a much higher power, but feedback errors can be accounted for within the \gls{mdp} formulation.

\subsection{Energy Model}
We assume that, on average, transmitting and receiving a single symbol requires energy of $P_s$ and $P_f$, respectively. The precise values of $P_s$ and $P_f$ depend on the physical implementation of the communication system and encompasses energy consumption of active \gls{rf} chains as well as encoding and decoding operations.
We parameterize $P_f=\alpha P_s$, where $\alpha \in [0, 1]$ models the relative energy expenditure of receiving feedback compared to transmitting (we assume that $P_f \le P_s$, as in most practical systems).

With this in mind, the average energy spent by the device in the $t$-th round is $E(t) = P_s(\hat{L}_p+\hat{L}_t) + \alpha P_s(\hat{L}_p+\hat{L}_{\mathrm{f}})$, and the average cumulative energy consumed up until and including the $t$-th round is $E_{\text{tot}}(t) = \sum_{i=0}^t E(i)$. We further denote the number of successful information bits transmitted per energy unit as $E_b(t) = \frac{K}{E_{\text{tot}}(t)}\mathds{1}[\hat{\mathbf{c}}_t=\mathbf{c}]$, where $\mathds{1}[\cdot]$ is the indicator function, equal to $1$ if the condition is true, and $0$ otherwise, and $\hat{\mathbf{c}}_t$ is the decoder's estimate of $\mathbf{c}$.

\subsection{HARQ Scheme}\label{sec:harq}
In the first round, $t=0$, message $m$ is encoded into an initial codeword $\mathbf{c}\in\code$ of $L_0$ code symbols, which is mapped to $\hat{L}_0$ channel symbols, and transmitted along with the preamble. In subsequent rounds, the transmitter constructs \emph{additional} $L_0$ code symbols, punctures them to $L_t\leq L_0$ symbols, and transmits the $\hat{L}_t$ channel symbols along with the preamble. The puncturing pattern is denoted $\mathbf{f}_t\in\{0, 1\}^{L_0}$ with a $1$ in the $i$-th entry indicating the transmission of the $i$-th symbol, such that $L_t=\sum_{i=1}^{L_0} \mathbf{f}_t(i)\leq L_0$. This continues until either the receiver is able to decode the message and the device receives an ACK, or a maximum of $T_{\mathrm{max}}$ rounds are completed.

\begin{figure*}[t]
    \centering
    \resizebox{0.7\linewidth}{!}{%
        \input{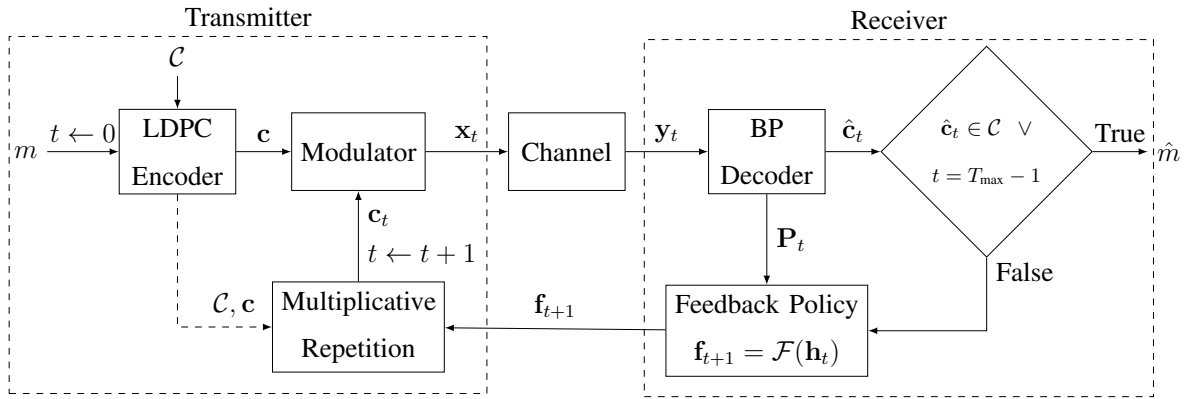}
    }%
    \caption{Block diagram of the proposed rich feedback protocol, \gls{raf}.}\vspace{-0.4cm}
    \label{fig:system_block_diagram}
\end{figure*}

The decoder employs \gls{bp} decoding with at most $I$ \gls{bp} iterations to obtain an estimate $\hat{\mathbf{c}}_t$ of $\mathbf{c}$ based on the received signals $\mathbf{y}_0,\ldots,\mathbf{y}_t$. The \gls{bp} decoder takes the \gls{snr} as input, and for simplicity, we will assume that the \gls{snr} is perfectly estimated using the preamble symbols. However, note that \gls{bp} decoders, such as the one used in this work, generally perform well even with imperfect estimates~\cite{Yuan2012Imperfect}. If $\hat{\mathbf{c}}_t$ is a valid codeword, i.e., $\hat{\mathbf{c}}_t\in\mathcal{C}$, the message is assumed to be correctly received and the receiver terminates the transmission by sending an ACK. On the other hand, if $\hat{\mathbf{c}}_t$ is not a valid codeword, then the receiver transmits a NACK and a feedback signal (unless $t=T_{\mathrm{max}}-1$, in which case the transmission is abandoned and the message is dropped).

The outlined \gls{harq} scheme can be employed with any rate adaptive code allowing for soft decoding, including polar codes, turbo codes, and more. However, in this work, we limit our attention to \gls{mr} non-binary \gls{ldpc} codes~\cite{DM98}, as these have proven to be efficient for short packet transmissions~\cite{Liva2020:ShortCodes}.

\section{Reinforcement-based Adaptive Feedback}\label{sec:soft_decoding}
In this section, we present our proposed adaptive feedback scheme. We first introduce the general feedback design, and then show how to learn the feedback policy using \gls{drl}. We propose to control the transmissions by providing a puncturing pattern in the feedback link. Specifically, in round $t$ we provide $\mathbf{f}_{t+1}\in \{0, 1\}^{L_0}$, indicating the binary puncturing pattern that should be used for the uplink transmission in round $t+1$. As a result, the uncoded feedback message has length $L_{0}$, and consequently $L_{\mathrm{f}}=\lceil\frac{L_0}{\log_2q}\rceil$ \footnote{For simplicity, we assume no encoding or compression of the puncturing pattern, noting that the flexibility in the energy model keeps the presentation general.}.

The objective is to learn a policy that chooses the optimal puncturing pattern by using the information from previous rounds, represented by the a posteriori probabilities estimated by the \gls{bp} decoder, which are denoted by the matrix $\mathbf{P}_t=\left[\text{Pr}\left(s_i=\zeta_j\mid \mathbf{y}_t,\dots,\mathbf{y}_0\right)\right]_{ij}$ where $\zeta_j$ is the $j$-th element in the \gls{gf} $\mathbb{F}_q$. We make the simplification to only consider the a posteriori entropy vector, $\mathbf{h}_t\in\mathcal{S}=[0,\log_2q]^{L_0}$, of the code symbols $s_i$, $i=1,2,\ldots,L_0$, of the code $\code$, defined as
\begin{equation}\label{eq:H}
    \mathbf{h}_t(i)=-\sum_{j=0}^{q-1} [\mathbf{P}_t]_{ij} \log_2[\mathbf{P}_t]_{ij}.
\end{equation}
The choice of using $\mathbf{h}_t$ as the state, as opposed to the full code symbol probability matrix $\mathbf{P}_t$, is justified by empirical results indicating that using $\mathbf{P}_t$ does not lead to improvements in the results. The feedback message in round $t$ is then constructed as $\mathbf{f}_{t+1}=\mathcal{F}(\mathbf{h}_t)$, where $\mathcal{F}:\mathcal{S}\to \{0,1\}^{L_0}$ is a deterministic function that maps the entropy vector to a message of $L_0$ bits indicating the puncturing pattern to be used in round $t+1$.
We formulate the objective as
\begin{equation}\label{eq:overall_opt}
    \mathcal{F}^*=\argmax_{\mathcal{F}:\mathcal{S}\rightarrow \{0,1\}^{L_0}} \E_{\mathcal{F}}\left[E_b(T)\gamma^T\right],
\end{equation}
where $\E_{\mathcal{F}}$ is the expected value operator given that the agent follows policy $\mathcal{F}$, $T$ is the number of transmission rounds until termination, and $\gamma \in (0, 1]$ is an exponential discount factor that is meant to ensure low latency (lower values of $\gamma$ lead the system to prioritize latency over energy efficiency). The expectation in Eq.~\eqref{eq:overall_opt} is over $\mathbf{c}$, generated from encoding a message $m$ drawn uniformly from the set of messages, and $T$ which given $\mathcal{F}$ and $\mathbf{c}$ depends only on the channel realizations.
For simplicity, we restrict our attention to stationary, deterministic policies.
The proposed rich feedback communication protocol is summarized in Fig.~\ref{fig:system_block_diagram}.

\subsection{MDP Formulation}
The problem introduced in Eq.~\eqref{eq:overall_opt} is approximated by an \gls{mdp} assuming that the next entropy vector is approximately independent of the past given the chosen action and the current entropy vector.
\footnote{This is only an approximation, since the posterior probabilities do not exactly satisfy the Markov property even though the channel realizations are independent.}
Under this assumption, the \gls{mdp} policy amounts to repeatedly applying $\mathcal{F}$ to the current state $\mathbf{h}_t$, followed by receiving a \emph{reward} $r_t$. The optimal function $\mathcal{F}^*$ can be found using \gls{rl} techniques, such as policy evaluation, policy iteration, value iteration, Q-learning, etc.~\cite{Sutton2018:Reinforcement}. In this work, we consider Q-learning with the purpose of exploiting the well-known \gls{dqn} approach~\cite{mnih2015human}.

The number of actions in the proposed \gls{mdp} is $2^{L_0}$.
To simplify, we instead consider selecting the \emph{number} of symbols to be retransmitted in the next round, and denote this policy as $\tilde{\mathcal{F}}:\mathcal{S}\to \mathcal{A}$, where $\mathcal{A}=\{1, \dots, L_0\}$ is the action space of the \gls{mdp} (the action of requesting zero symbols is selected automatically if $\hat{\mathbf{c}}_t\in\code$).
The symbols to be retransmitted are then requested deterministically as the ones with the highest entropy computed using Eq.~\eqref{eq:H}, i.e.,
\begin{equation}
  \big[\mathcal{F}(\mathbf{h}_t))\big]_i = \mathds{1}\Bigg[\sum_{j\neq i} \mathds{1}[\mathbf{h}_t(i) < \mathbf{h}_t(j)] < \tilde{\mathcal{F}}(\mathbf{h}_t)\Bigg],
\end{equation}
where $\tilde{\mathcal{F}}(\mathbf{h}_t)$ is the number of symbols to be transmitted.
The simplification above is justified by empirical results that suggest optimizing an agent to select which symbols to retransmit yields no performance gains compared to selecting the maximum entropy symbols.

Due to the episodic nature of the \gls{mdp}, we set rewards $r_t=0$ for $t = 1, \dots, T-1$, and only assign non-zero rewards at the end of a transmission (episode).
With this structure, the final \gls{mdp} is defined as finding $\tilde{\mathcal{F}}^*$ that solves
\begin{equation}\label{eq:mdp_opt}
    \tilde{\mathcal{F}}^*=\argmax_{\tilde{\mathcal{F}}:\mathcal{S}\rightarrow \mathcal{A}}~\E_{\mathcal{F}}\left[r_T\gamma^T\right],
\end{equation}
where $r_T$ is the reward for the end of the episode. While $E_b(T)$ would be a natural choice for rewards,
we have empirically observed faster convergence and more robust training across simulation settings by defining the reward as
\begin{equation}\label{eq:reward_scheme_orig}
    r_T = (2\cdot\mathds{1}[\hat{\mathbf{c}}_T=\mathbf{c}] - 1) - \frac{E_{\text{tot}}(T)}{\E_{\mathcal{F}_{0}}[E_{\text{tot}}(T)]},
\end{equation}
where $\mathcal{F}_{0}$ is
a naïve static policy taking actions $L_t=1,\ \forall t>0$,
and is included as it provided improvements in convergence of the learning algorithm. We have that, $\E_{\mathcal{F}_{0}}[E_{\text{tot}}(T)] = \E_{\mathcal{F}_{0}}[T] P_s(\hat{L}_p(1+\alpha)+\alpha\hat{L}_{\mathrm{f}}+1) + P_s(\hat{L}_p(1+\alpha)+\alpha\hat{L}_{\mathrm{f}}+\hat{L}_0)$. Now, by simulating episodes using the naïve static policy, we can approximate $\E_{\mathcal{F}_{0}}[T]$ by the average number of rounds, and in turn approximate $\E_{\mathcal{F}_{0}}[E_{\text{tot}}(T)]$. 

As mentioned, we will use Q-learning to solve this \gls{mdp}. In Q-learning, the objective is to use observations of the environment to learn an \emph{action-value} function, $Q(\mathbf{h},a)$, that characterizes
the expected long-term reward of taking action $a$ in state $\mathbf{h}$.
Given the estimated action-value function and a state, $\mathbf{h}_t$, the action is selected as $a_t = \argmax_{a\in\mathcal{A}}~Q(\mathbf{h}_t, a)$.

The state transitions in the \gls{mdp} are complex, as they depend both on the transmission of the new code symbols and the \gls{bp} decoder, so we cannot model them analytically. Additionally, since $\mathcal{S} = [0,\log_2q]^{L_0}$, the state space is infinite, so using a lookup table would not be feasible~\cite{Baird1995:Residual}. Instead, we consider the \gls{dqn} approach~\cite{mnih2015human}, which uses a neural network to approximate the value function.

\subsection{Deep Q-Learning}
The \gls{dqn} takes the entropy vector $\mathbf{h}_t$ as input, and returns an estimate of $Q(\mathbf{h}_t, a_t)$, $\forall a_t \in \mathcal{A}$. We employ a fully connected deep neural network using \gls{relu} activation functions with number of layers and number of hidden units in the respective layers expressed by the tuple $N_{\text{hidden}}$, and $L_0$ output units.

For training, we adopt replay memory~\cite{mnih2015human} where agent experience is stored in tuples $e_t=(\mathbf{h}_t, a_t, r_t, \mathbf{h}_{t+1})$. The replay memory buffer size is denoted by $\text{MEM}$, and as new experience samples are observed the oldest experience is deleted from the replay memory buffer. During training, batches of size $\text{BS}$ are sampled from the replay memory in order to optimize the neural network parameters using backpropagation with the ADAM algorithm and gradient clipping with max gradient norm, GC. The number of episodes between training sessions is $F_{\text{upd}}$ and the number of parameter updating steps per training session is $N_{\text{upd}}$.

Moreover, as in~\cite{mnih2015human}, to avoid biases, two different $Q$ functions, denoted by $Q_A$ and $Q_B$, are used to select and evaluate actions, respectively. The model is updated by
\begin{equation}\label{eq:RL_update}
    \begin{split}
        Q_A^{\text{new}}(\mathbf{h}_t, a_t) &= (1 - \eta) Q_A(\mathbf{h}_t, a_t) \\&\quad+ \eta \big(r_t + \gamma_{_{\text{DQN}}} \max_{a_{t+1}\in \mathcal{A}} Q_B(\mathbf{h}_{t+1}, a_{t+1})\big),
    \end{split}
\end{equation}
where $\eta$ is the learning rate and $\gamma_{_{\text{DQN}}}$ is the \gls{dqn} discount factor. When $\gamma_{_{\text{DQN}}}=0$, the agent will strive to achieve immediate rewards while when $\gamma_{_{\text{DQN}}}=1$, the agent evaluates the actions based on the sum total of the future rewards. The parameters of $Q_B$ are then updated by copying the parameters of $Q_A$ after every $F_{\text{mod}}$ training sessions.

\setlength{\tabcolsep}{0.2em} 
\begin{table}[t]
    \renewcommand{\arraystretch}{0.95}
    \centering
    \notsotiny
    \caption{Scenario and \gls{dqn} settings}\label{tab:params}
    \begin{tabular}{ccc|ccc}
        \toprule
        \textbf{Parameter}	& \textbf{Value} & \textbf{Description} & \textbf{Parameter}	& \textbf{Value} & \textbf{Description} \\ \midrule
        $K$ &40~bits & Payload size&        $L_0$ & 15~symbols & Codeword length\\
        $q$ & 256 & Galois field order&        SNR & 6~dB & Received SNR\\
        $I$ & 5 & Max \gls{bp} iterations&        $E_{\text{first}}$ & 0.6~mJ & Energy reference~\cite{bouguera2018energy}\\
        $\tau_p$ & 1~ms & Packet slot duration&        $\E_{\mathcal{F}_0}[T]$ & 7.7 & Normalization factor\\
        $T_{\mathrm{max}}$ & 15 & Max rounds&        $\eta$ & 0.001 & Learning rate\\
        BS & 64 & Batch size&        $F_{\text{upd}}$ & 100 & Training frequency\\
        GC & 5 & Max gradient norm &        MEM & 60000 & Memory replay buffer\\
        $N_{\text{upd}}$ & 15 & Training steps per update &        $N_{\text{hidden}}$ & [64, 32, 16] & Hidden layer sizes\\
        $F_{\text{mod}}$ & 10 & Target update frequency&         $\varepsilon_d$ & 0.05 & Exploration decay rate\\
        $\varepsilon_0$ & 1 & Initial exploration rate &         $F_\varepsilon$ & 8000 & Exploration decay freq.\\
        $\varepsilon_{\text{min}}$ & 0.4 & Min exploration rate &         $N_{\text{test}}$ & 300000 & Test episodes\\
        $N_{\text{train}}$ & 270000 & Training episodes &         \\
        \bottomrule
    \end{tabular}
    \vspace{-0.5cm}
\end{table}

During experience gathering, the agent uses the $\varepsilon$-greedy policy. The $\varepsilon$-greedy parameter is initialized as $\varepsilon_0$ and decreases linearly with a decay parameter $\varepsilon_d$ every $F_\varepsilon$ simulated episodes, with a minimum preset as $\varepsilon_{\text{min}}$.

\section{Simulation Settings and Results}\label{sec:sim}
We define a scenario corresponding to the one given in Sec.~\ref{sec:soft_decoding}, with a transmitter encoding packets containing $K=40$~bits of information into $L_0=15$ symbols using a codebook of order $q=256$, i.e., with code rate $1/3$.
The transmitter employs \gls{mr} non-binary \gls{ldpc} codes~\cite{Kasai2011:MRLDPC}, defined over $\mathbb{F}_q$, which allows the device to generate codes with successively lower rates by concatenating previous codewords with a new set of symbols.
In the following, we will consider a modulation order of $M=256$ unless otherwise stated. Although this setting is more suitable for complex applications than for \gls{iot}~\cite{Kumar2021:QAM}, it simplifies the presentation: with this value, each code symbol maps exactly to one constellation point, i.e., $\hat{L}_t=L_t$. We will also show results for \gls{qpsk} ($M=4$) below. We consider {four} baseline policies as benchmarks:
\begin{enumerate}
    \item \emph{\gls{harq}}: we fix $L_t$ to a constant for each $t>0$ and use a random uniform symbol selection policy to select $L_t$ different symbols. This policy only requires 1~bit of feedback as with traditional \gls{harq}. We set $L_{\mathrm{f}}=1$.
    \item \emph{\gls{dharq}}: we adaptively determine $L_t$ as the number of symbols that have an entropy higher than a given threshold value $H_{\text{th}}$. This is an adaptation of the ideas in~\cite{Visotsky2005:RBHARQ} to the present scenario. It requires 4~bits of feedback and we set $L_{\mathrm{f}}=1$.
    \item \emph{\Gls{st} \gls{harq}}: we fix $L_t$ to a constant for each $t>0$. Unlike the \gls{harq} and \gls{dharq} policies, this policy selects the maximum entropy symbols, and is an adaptation of the ideas in~\cite{Shea2002:Reliability} to the present scenario.
    The full 15~bit feedback is required, and we set $L_{\mathrm{f}}=2$.
    \item \emph{\Gls{ta} \gls{harq}}: this policy selects for retransmission all the symbols that have an entropy higher than a given threshold value $H_{\text{th}}$, and is based on the ideas in~\cite{Roongta2003:RBHARQ}. It requires the full 15~bit feedback. Again, we set $L_{\mathrm{f}}=2$.
\end{enumerate}

In this study, for $L_p = 0$ the energy cost of the first round is fixed as $E_{\text{first}} = 0.6$~mJ, {and a constant packet transmission slot of $\tau_p = 1$~ms is set to avoid difficulties with collisions in multi-device communications~\cite{szczecinski2013rate}}.
These values are consistent with empirical measurements for LoRa deployments~\cite{bouguera2018energy}.
All parameters for the \gls{dqn} architecture and training, as well as the simulation scenario, are given in Table~\ref{tab:params}.

Given the transmission protocol specification $L_p$, the hardware specification $\alpha$, and the desired energy-latency trade-off through the exponential discount factor $\gamma$, we can optimize hyperparameters for the baseline policies and \gls{raf}.

Consider the optimization of the baseline policies by grid search: For the \gls{dharq} and \gls{ta} policies, the only parameter is $H_{\text{th}}$ while for the static policies the parameters are $L_t\in\mathcal{A}$ for $t=1,\dots,T_{\mathrm{max}}-1$. To make a grid search feasible,
the sequence $L_t$ is defined in terms of a single parameter $L_{\text{static}}\in\{1, \dots, L_0\}$ as $L_t=L_{\text{static}}$ if $t=1$, $L_t=L_{\text{static}}\ \forall t>1$ if $1 \leq L_{\text{static}} \leq 2$, $L_t=2,\ \forall t>1$ if $2 < L_{\text{static}} \leq 10$ and $L_t=4,\ \forall t>1$ if $L_{\text{static}} > 10$.
This definition is motivated by empirical results indicating that the optimal policy is tapered.

\begin{figure}[t]
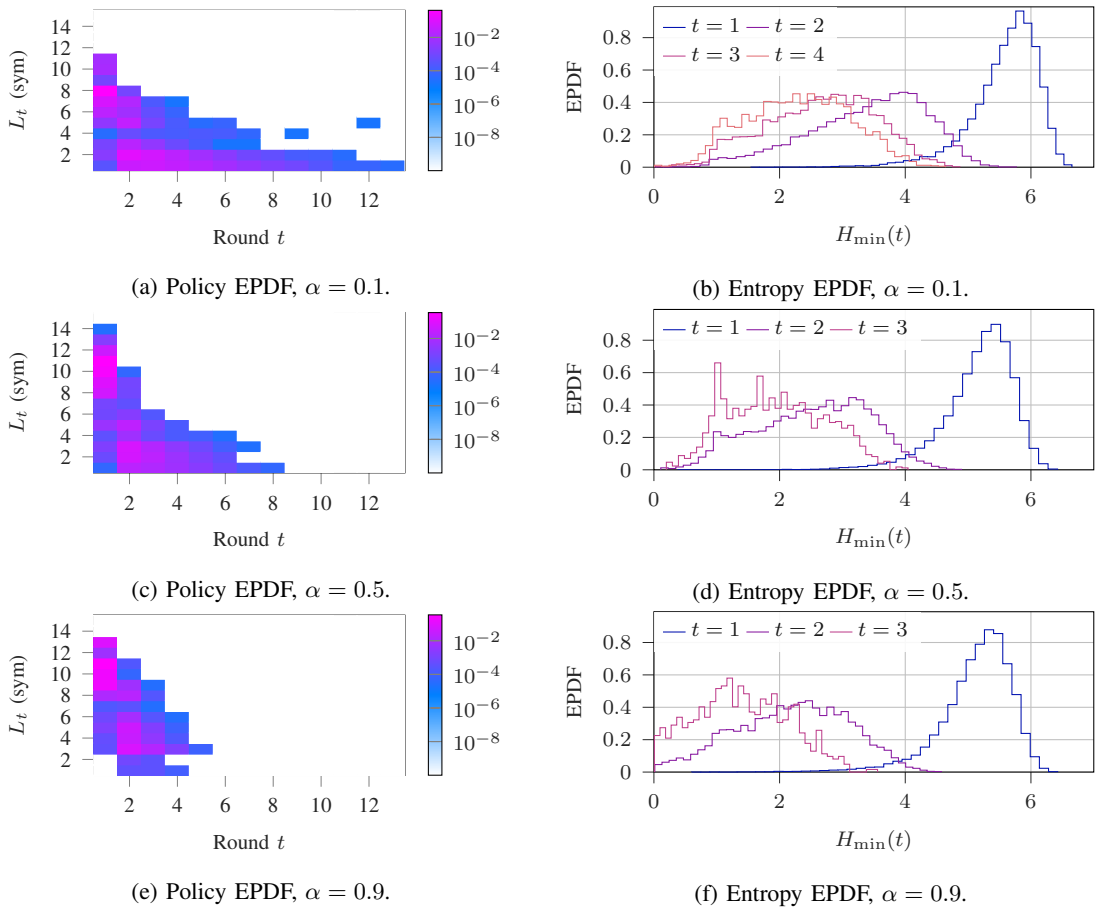

    \centering
    \begin{minipage}{0.24\textwidth}
        \centering
        \subfloat[Policy EPDF, $\alpha=0.1$.]{\begin{tikzpicture}
 \begin{axis}[view={0}{90},
        xlabel={Round $t$},
        xlabel style={color=white!15!black,font=\scriptsize},
        ylabel style={color=white!15!black,font=\scriptsize},
        ylabel={$L_t$ (sym)},
        xticklabel style={color=white!15!black,font=\scriptsize},
        yticklabel style={color=white!15!black,font=\scriptsize},
        ytick={2, 4, 6, 8, 10, 12, 14},
        yticklabels={2, 4, 6, 8, 10, 12, 14},
        xtick={2, 4, 6, 8, 10, 12},
        xticklabels={2, 4, 6, 8, 10, 12},
        xmin=0.5,
        xmax=13.5,
        ymin=0.5,
        ymax=15.5,
        colormap/cool,
        tick align=outside,
        tick pos=left,
        colorbar,
        colorbar/width=1.8mm,
        width=0.9\gfwidth,
        height=0.8\gfwidth,
        colorbar style={
            ytick={-8,-6,-4,-2,0},
            yticklabel={$10^{\pgfmathprintnumber{\tick}}$},
            ylabel style={color=white!15!black,font=\scriptsize},
            yticklabel style={color=white!15!black,font=\scriptsize},
        }
        ]
    \addplot[matrix plot*,point meta=explicit] table [
                    row sep=newline,
                    x=x,
                    y=y,
                    meta=logz,]
        {
        x y logz
        1 1 -3.504702680204397
        1 2 -2.6178803059269065
        1 3 -2.909076798881844
        1 4 -4.43412160591869
        1 5 -2.4563980006298416
        1 6 -1.638589163208535
        1 7 -1.1389945206664982
        1 8 -0.3740851518287283
        1 9 -2.9358110521290888
        1 10 -2.002757841759702
        1 11 -2.063053743646953
        1 12 -10.0
        1 13 -10.0
        1 14 -10.0
        1 15 -10.0
        
        2 1 -1.3814276639937215
        2 2 -0.8944183669708639
        2 3 -1.6910039807039479
        2 4 -3.216637661704783
        2 5 -1.5181943942215737
        2 6 -2.1341787058959225
        2 7 -1.8298955528342193
        2 8 -2.9642995899405262
        2 9 -10.0
        2 10 -10.0
        2 11 -10.0
        2 12 -10.0
        2 13 -10.0
        2 14 -10.0
        2 15 -10.0
        
        3 1 -1.3284411429728806
        3 2 -1.2287818844871665
        3 3 -2.431955544162182
        3 4 -3.7351516015826705
        3 5 -2.909076798881844
        3 6 -3.303787837423683
        3 7 -3.7351516015826705
        3 8 -10.0
        3 9 -10.0
        3 10 -10.0
        3 11 -10.0
        3 12 -10.0
        3 13 -10.0
        3 14 -10.0
        3 15 -10.0
        
        4 1 -1.5142593523631513
        4 2 -1.6592398400998931
        4 3 -3.1553680049658603
        4 4 -3.832061614590727
        4 5 -3.7809090921433457
        4 6 -4.133091610254708
        4 7 -4.73515160158267
        4 8 -10.0
        4 9 -10.0
        4 10 -10.0
        4 11 -10.0
        4 12 -10.0
        4 13 -10.0
        4 14 -10.0
        4 15 -10.0
        
        5 1 -1.7551482299989243
        5 2 -2.1385545059562103
        5 3 -3.832061614590727
        5 4 -3.7351516015826705
        5 5 -4.43412160591869
        5 6 -10.0
        5 7 -10.0
        5 8 -10.0
        5 9 -10.0
        5 10 -10.0
        5 11 -10.0
        5 12 -10.0
        5 13 -10.0
        5 14 -10.0
        5 15 -10.0
        
        6 1 -2.0761867589182357
        6 2 -2.5678342668344944
        6 3 -4.73515160158267
        6 4 -3.832061614590727
        6 5 -4.133091610254708
        6 6 -10.0
        6 7 -10.0
        6 8 -10.0
        6 9 -10.0
        6 10 -10.0
        6 11 -10.0
        6 12 -10.0
        6 13 -10.0
        6 14 -10.0
        6 15 -10.0
        
        7 1 -2.379125744389548
        7 2 -3.0539103642070833
        7 3 -4.73515160158267
        7 4 -4.258030346863008
        7 5 -10.0
        7 6 -10.0
        7 7 -10.0
        7 8 -10.0
        7 9 -10.0
        7 10 -10.0
        7 11 -10.0
        7 12 -10.0
        7 13 -10.0
        7 14 -10.0
        7 15 -10.0
        
        8 1 -2.7666686530287357
        8 2 -3.337211592910633
        8 3 -10.0
        8 4 -10.0
        8 5 -10.0
        8 6 -10.0
        8 7 -10.0
        8 8 -10.0
        8 9 -10.0
        8 10 -10.0
        8 11 -10.0
        8 12 -10.0
        8 13 -10.0
        8 14 -10.0
        8 15 -10.0
        
        9 1 -3.0361815972466517
        9 2 -3.832061614590727
        9 3 -10.0
        9 4 -4.73515160158267
        9 5 -10.0
        9 6 -10.0
        9 7 -10.0
        9 8 -10.0
        9 9 -10.0
        9 10 -10.0
        9 11 -10.0
        9 12 -10.0
        9 13 -10.0
        9 14 -10.0
        9 15 -10.0
        
        10 1 -3.559060342526989
        10 2 -3.8900535615684135
        10 3 -10.0
        10 4 -10.0
        10 5 -10.0
        10 6 -10.0
        10 7 -10.0
        10 8 -10.0
        10 9 -10.0
        10 10 -10.0
        10 11 -10.0
        10 12 -10.0
        10 13 -10.0
        10 14 -10.0
        10 15 -10.0
        
        11 1 -4.133091610254708
        11 2 -4.43412160591869
        11 3 -10.0
        11 4 -10.0
        11 5 -10.0
        11 6 -10.0
        11 7 -10.0
        11 8 -10.0
        11 9 -10.0
        11 10 -10.0
        11 11 -10.0
        11 12 -10.0
        11 13 -10.0
        11 14 -10.0
        11 15 -10.0
        
        12 1 -4.133091610254708
        12 2 -10.0
        12 3 -10.0
        12 4 -10.0
        12 5 -4.73515160158267
        12 6 -10.0
        12 7 -10.0
        12 8 -10.0
        12 9 -10.0
        12 10 -10.0
        12 11 -10.0
        12 12 -10.0
        12 13 -10.0
        12 14 -10.0
        12 15 -10.0
        
        13 1 -4.43412160591869
        13 2 -10.0
        13 3 -10.0
        13 4 -10.0
        13 5 -10.0
        13 6 -10.0
        13 7 -10.0
        13 8 -10.0
        13 9 -10.0
        13 10 -10.0
        13 11 -10.0
        13 12 -10.0
        13 13 -10.0
        13 14 -10.0
        13 15 -10.0
        
        14 1 -10.0
        14 2 -10.0
        14 3 -10.0
        14 4 -10.0
        14 5 -10.0
        14 6 -10.0
        14 7 -10.0
        14 8 -10.0
        14 9 -10.0
        14 10 -10.0
        14 11 -10.0
        14 12 -10.0
        14 13 -10.0
        14 14 -10.0
        14 15 -10.0
        
        15 1 -10.0
        15 2 -10.0
        15 3 -10.0
        15 4 -10.0
        15 5 -10.0
        15 6 -10.0
        15 7 -10.0
        15 8 -10.0
        15 9 -10.0
        15 10 -10.0
        15 11 -10.0
        15 12 -10.0
        15 13 -10.0
        15 14 -10.0
        15 15 -10.0
};
\end{axis}
\end{tikzpicture}\label{subfig:alpha01_Lp1_matrix}}
    \end{minipage}
    \begin{minipage}{0.24\textwidth}
        \centering
        \subfloat[Entropy EPDF, $\alpha=0.1$.]{\input{tikzplotlib_figs_icc/alpha01_Lp1}\label{subfig:alpha01_Lp1}}
    \end{minipage}\\
    \begin{minipage}{0.24\textwidth}
        \centering
        \subfloat[Policy EPDF, $\alpha=0.5$.]{\begin{tikzpicture}
 \begin{axis}[view={0}{90},
        xlabel={Round $t$},
        xlabel style={color=white!15!black,font=\scriptsize},
        ylabel style={color=white!15!black,font=\scriptsize},
        ylabel={$L_t$ (sym)},
        xticklabel style={color=white!15!black,font=\scriptsize},
        yticklabel style={color=white!15!black,font=\scriptsize},
        ytick={2, 4, 6, 8, 10, 12, 14},
        yticklabels={2, 4, 6, 8, 10, 12, 14},
        xtick={2, 4, 6, 8, 10, 12},
        xticklabels={2, 4, 6, 8, 10, 12},
        xmin=0.5,
        xmax=13.5,
        ymin=0.5,
        ymax=15.5,
        colormap/cool,
        tick align=outside,
        tick pos=left,
        colorbar,
        colorbar/width=1.8mm,
        width=0.9\gfwidth,
        height=0.8\gfwidth,
        colorbar style={
            ytick={-8,-6,-4,-2,0},
            yticklabel={$10^{\pgfmathprintnumber{\tick}}$},
            ylabel style={color=white!15!black,font=\scriptsize},
            yticklabel style={color=white!15!black,font=\scriptsize},
        }
        ]
    \addplot[matrix plot*,point meta=explicit] table [
                    row sep=newline,
                    x=x,
                    y=y,
                    meta=logz,]
        {
        x y logz
        1 1 -4.2829957579913245
        1 2 -3.222297917637713
        1 3 -2.8205977600923684
        1 4 -3.437897717977068
        1 5 -2.351029643263152
        1 6 -3.1069044989356436
        1 7 -3.0277232528880185
        1 8 -1.3392539921600108
        1 9 -1.0684168044208255
        1 10 -0.6057082569085547
        1 11 -0.46118755956603963
        1 12 -1.4229573681841312
        1 13 -2.7147940339243295
        1 14 -4.584025753655306
        1 15 -10.0
        
        2 1 -1.7439196591985482
        2 2 -1.2829957579913247
        2 3 -1.223621698925367
        2 4 -2.247566019806776
        2 5 -1.6654712231050324
        2 6 -2.7089644902636056
        2 7 -3.437897717977068
        2 8 -3.092664059821033
        2 9 -3.052546836613051
        2 10 -4.2829957579913245
        2 11 -10.0
        2 12 -10.0
        2 13 -10.0
        2 14 -10.0
        2 15 -10.0
        
        3 1 -1.9308132398799622
        3 2 -1.7055039581540994
        3 3 -1.885925208031916
        3 4 -3.2416030728330996
        3 5 -2.9308132398799622
        3 6 -3.6809357666633624
        3 7 -10.0
        3 8 -10.0
        3 9 -10.0
        3 10 -10.0
        3 11 -10.0
        3 12 -10.0
        3 13 -10.0
        3 14 -10.0
        3 15 -10.0
        
        4 1 -2.2576898927265545
        4 2 -2.3029923864075785
        4 3 -2.5711885289501337
        4 4 -3.8058745032716623
        4 5 -3.8058745032716623
        4 6 -10.0
        4 7 -10.0
        4 8 -10.0
        4 9 -10.0
        4 10 -10.0
        4 11 -10.0
        4 12 -10.0
        4 13 -10.0
        4 14 -10.0
        4 15 -10.0
        
        5 1 -2.670211901271589
        5 2 -2.8938296736267923
        5 3 -3.2038145119437
        5 4 -3.9819657623273432
        5 5 -10.0
        5 6 -10.0
        5 7 -10.0
        5 8 -10.0
        5 9 -10.0
        5 10 -10.0
        5 11 -10.0
        5 12 -10.0
        5 13 -10.0
        5 14 -10.0
        5 15 -10.0
        
        6 1 -3.1860857449832682
        6 2 -3.407934494599625
        6 3 -3.9819657623273432
        6 4 -4.584025753655306
        6 5 -10.0
        6 6 -10.0
        6 7 -10.0
        6 8 -10.0
        6 9 -10.0
        6 10 -10.0
        6 11 -10.0
        6 12 -10.0
        6 13 -10.0
        6 14 -10.0
        6 15 -10.0
        
        7 1 -3.8058745032716623
        7 2 -10.0
        7 3 -4.584025753655306
        7 4 -10.0
        7 5 -10.0
        7 6 -10.0
        7 7 -10.0
        7 8 -10.0
        7 9 -10.0
        7 10 -10.0
        7 11 -10.0
        7 12 -10.0
        7 13 -10.0
        7 14 -10.0
        7 15 -10.0
        
        8 1 -4.2829957579913245
        8 2 -10.0
        8 3 -10.0
        8 4 -10.0
        8 5 -10.0
        8 6 -10.0
        8 7 -10.0
        8 8 -10.0
        8 9 -10.0
        8 10 -10.0
        8 11 -10.0
        8 12 -10.0
        8 13 -10.0
        8 14 -10.0
        8 15 -10.0
        
        9 1 -10.0
        9 2 -10.0
        9 3 -10.0
        9 4 -10.0
        9 5 -10.0
        9 6 -10.0
        9 7 -10.0
        9 8 -10.0
        9 9 -10.0
        9 10 -10.0
        9 11 -10.0
        9 12 -10.0
        9 13 -10.0
        9 14 -10.0
        9 15 -10.0
        
        10 1 -10.0
        10 2 -10.0
        10 3 -10.0
        10 4 -10.0
        10 5 -10.0
        10 6 -10.0
        10 7 -10.0
        10 8 -10.0
        10 9 -10.0
        10 10 -10.0
        10 11 -10.0
        10 12 -10.0
        10 13 -10.0
        10 14 -10.0
        10 15 -10.0
        
        11 1 -10.0
        11 2 -10.0
        11 3 -10.0
        11 4 -10.0
        11 5 -10.0
        11 6 -10.0
        11 7 -10.0
        11 8 -10.0
        11 9 -10.0
        11 10 -10.0
        11 11 -10.0
        11 12 -10.0
        11 13 -10.0
        11 14 -10.0
        11 15 -10.0
        
        12 1 -10.0
        12 2 -10.0
        12 3 -10.0
        12 4 -10.0
        12 5 -10.0
        12 6 -10.0
        12 7 -10.0
        12 8 -10.0
        12 9 -10.0
        12 10 -10.0
        12 11 -10.0
        12 12 -10.0
        12 13 -10.0
        12 14 -10.0
        12 15 -10.0
        
        13 1 -10.0
        13 2 -10.0
        13 3 -10.0
        13 4 -10.0
        13 5 -10.0
        13 6 -10.0
        13 7 -10.0
        13 8 -10.0
        13 9 -10.0
        13 10 -10.0
        13 11 -10.0
        13 12 -10.0
        13 13 -10.0
        13 14 -10.0
        13 15 -10.0
        
        14 1 -10.0
        14 2 -10.0
        14 3 -10.0
        14 4 -10.0
        14 5 -10.0
        14 6 -10.0
        14 7 -10.0
        14 8 -10.0
        14 9 -10.0
        14 10 -10.0
        14 11 -10.0
        14 12 -10.0
        14 13 -10.0
        14 14 -10.0
        14 15 -10.0
        
        15 1 -10.0
        15 2 -10.0
        15 3 -10.0
        15 4 -10.0
        15 5 -10.0
        15 6 -10.0
        15 7 -10.0
        15 8 -10.0
        15 9 -10.0
        15 10 -10.0
        15 11 -10.0
        15 12 -10.0
        15 13 -10.0
        15 14 -10.0
        15 15 -10.0
};
\end{axis}
\end{tikzpicture}\label{subfig:alpha05_Lp1_matrix}}
    \end{minipage}
    \begin{minipage}{0.24\textwidth}
        \centering
        \subfloat[Entropy EPDF, $\alpha=0.5$.]{\input{tikzplotlib_figs_icc/alpha05_Lp1}\label{subfig:alpha05_Lp1}}
    \end{minipage}
    \caption{Colormaps showing the EPDF of chosen actions (left) and $H_{\min}$ (right) with \gls{raf} for $L_p=1$ and $\gamma_{_{\text{DQN}}}=0.5$.}\vspace{-0.4cm}
    \label{fig:alpha_raf}
\end{figure}

\subsection{Adaptivity of \Gls{raf}}
Fig.~\ref{fig:alpha_raf} provides intuition for the choices made by the \gls{raf} scheme. The figures on the left depict the \gls{epdf} of the chosen actions.
We can notice that the strategy is generally tapered, i.e., more symbols are retransmitted in the first few rounds, and that higher values of $\alpha$ correspond to fewer retransmitted symbols per round and more rounds. The figures on the right show the \gls{epdf} of the minimum entropy among the transmitted symbols in the first few rounds, i.e., the optimal entropy threshold, denoted $H_{\text{min}}(t)$.
Interestingly, the entropy distribution changes significantly from the first to the second round, but only slightly in subsequent rounds: this is due to the limited amount of information delivered in shorter retransmitted packets. Finally, we note that the distribution of the minimum entropy among the transmitted symbols has a high variance.
The optimal action is then hard to capture with a fixed threshold, making the \gls{ta} policy highly suboptimal.

\setlength{\tabcolsep}{0.2em} 
\begin{table}[t]
   \centering
   \caption{Comparison of \gls{raf} with the baseline methods}
   \scriptsize
   \label{tab:energy}
   \begin{tabular}{@{}r|ccc|ccc|ccc@{}}
   \toprule
   \multirow{2}{*}{\textbf{Scheme}} &\multicolumn{3}{c|}{$\mathbf{\E[T]}$ \textbf{(ms)}} & \multicolumn{3}{c|}{$\mathbf{\E[E_b(T)]}$ \textbf{(bits/mJ)}} & \multicolumn{3}{c}{{$\mathbf{\E[\mathds{1}[\hat{\mathbf{c}} \neq \mathbf{c}]]\cdot 10^{4}}$}} \\
           & $\alpha\shorteq0.1$ & $\alpha\shorteq0.5$ & $\alpha\shorteq0.9$ & $\alpha\shorteq0.1$ & $\alpha\shorteq0.5$ & $\alpha\shorteq0.9$ & {$\alpha\shorteq0.1$} & {$\alpha\shorteq0.5$} & {$\alpha\shorteq0.9$}  \\ \midrule
        HARQ    & 1.63    & 1.63    & 1.44    & 36.12  & \textbf{34.67}  & \textbf{33.12} & {1.79} & {1.79} & {1.24}   \\
        {D-HARQ}  & {1.80}    & {1.55}    & {1.38}    & {35.63}  & {33.79}  & {32.38} & {1.61} & {1.00} & {0.79}   \\
        ST      & 1.59    & 1.59    & \textbf{1.19}    & 35.90  & 33.42  & 30.89 & {0.97} & {0.97} & {0.39} \\
        TA      & 1.73    & 1.50    & 1.35    & 35.40  & 32.59  & 30.58 & {0.70} & {0.45} & {0.61} \\
        RAF     & \textbf{1.56}    & \textbf{1.39}    & 1.39    & \textbf{36.38}  & 33.58  & 31.65 & {\textbf{0.43}} & {\textbf{0.31}} & {\textbf{0.31}} \\
   \bottomrule
   \end{tabular}
   \vspace{-0.4cm}
\end{table}
\subsection{Comparison of Optimized Policies}

Let us consider the performance of \gls{raf} when $L_p=1$, trained with different values of $\alpha$: Table~\ref{tab:energy} shows the average performance for the various policies in terms of latency, successful bits per mJ, and \gls{uder}, with boldface indicating the best value in each column.
For each policy, $\mathbb{E}[E_b(T)\gamma^T]$ with $\gamma=0.92$ is maximized according to the respective parameters: $L_{\text{static}}$, $H_{\text{th}}$, and $\gamma_{_{\text{DQN}}}$.

We observe from Table~\ref{tab:energy} that \gls{ta} has the lowest number of successfully transmitted bits per energy unit, and while \gls{st} has a slightly lower latency than \gls{harq}, the cost of rich feedback makes \gls{harq} the preferred baseline. Looking at the \gls{uder}, we note that \gls{harq} and \gls{dharq} tend to have more undetected errors than their \gls{rbharq} counterparts in \gls{st} and \gls{ta}, and overall \gls{raf} has the lowest \gls{uder}. This highlights another advantage of \gls{raf}: retransmitting the highest-entropy symbols can steer the \gls{bp} decoder towards the right codeword, reducing undetected errors.

When $\alpha=0.1$, \gls{raf} is the most energy efficient policy while having a lower latency than the static baselines resulting in the best performance for the energy efficiency-latency trade-off. When comparing \gls{raf} with \gls{harq} with $\alpha=0.1$, we are able to reduce the error rate by a {factor four}, while also reducing the latency by $3.8$ percent and increasing the energy efficiency by $0.71$ percent. However, for higher $\alpha$-values, the cost of feedback outweighs the benefits of \gls{raf}.

\subsection{Pareto Dominance of RAF}
We compare \gls{raf} to the baselines in the latency and energy-efficiency trade-off using the \emph{Pareto frontier}
for $\alpha=0.1$, as \gls{raf} targets scenarios when receiving feedback consumes much less energy than transmitting.
For this comparison, different hyperparameter settings are tested: $L_{\text{static}}=1,2,\dots,15$ for \gls{harq} and \gls{st}, $H_{\text{th}}=2.9,3.1,\dots,7.1$ for \gls{dharq} and \gls{ta}, and $\gamma_{_{\text{DQN}}}=0.1,0.2,\dots,0.9$, $\alpha=0.1,0.5$, and $L_p=1,2$ for \gls{raf}.

Figs.~\ref{fig:pareto}a and \ref{fig:pareto}b depict the Pareto frontiers when $\alpha=0.1$ and $L_p=1,2$ for the average number of successfully transmitted bits per energy unit versus the average latency (in ms). These results show that \gls{raf} Pareto dominates the baselines for both preamble lengths, achieving a lower latency for the same energy efficiency (or, conversely, a higher energy efficiency with the same latency).
We show now that \gls{raf} can adapt to different modulation schemes by considering \gls{qpsk} modulation.
To be comparable with the $256$-\gls{qam} modulation, an order $256$ \gls{gf} value is mapped to four \gls{qpsk} symbols, each with normalized energy, and the \gls{snr} is lowered to $-3$ dB.
The results are shown in Figs.~\ref{fig:pareto}c and \ref{fig:pareto}d: the \gls{raf} model can adapt to the modulation scheme, yielding Pareto dominant results for both of the considered modulation schemes.
\def\gfheight{0.55\linewidth}
\begin{figure}
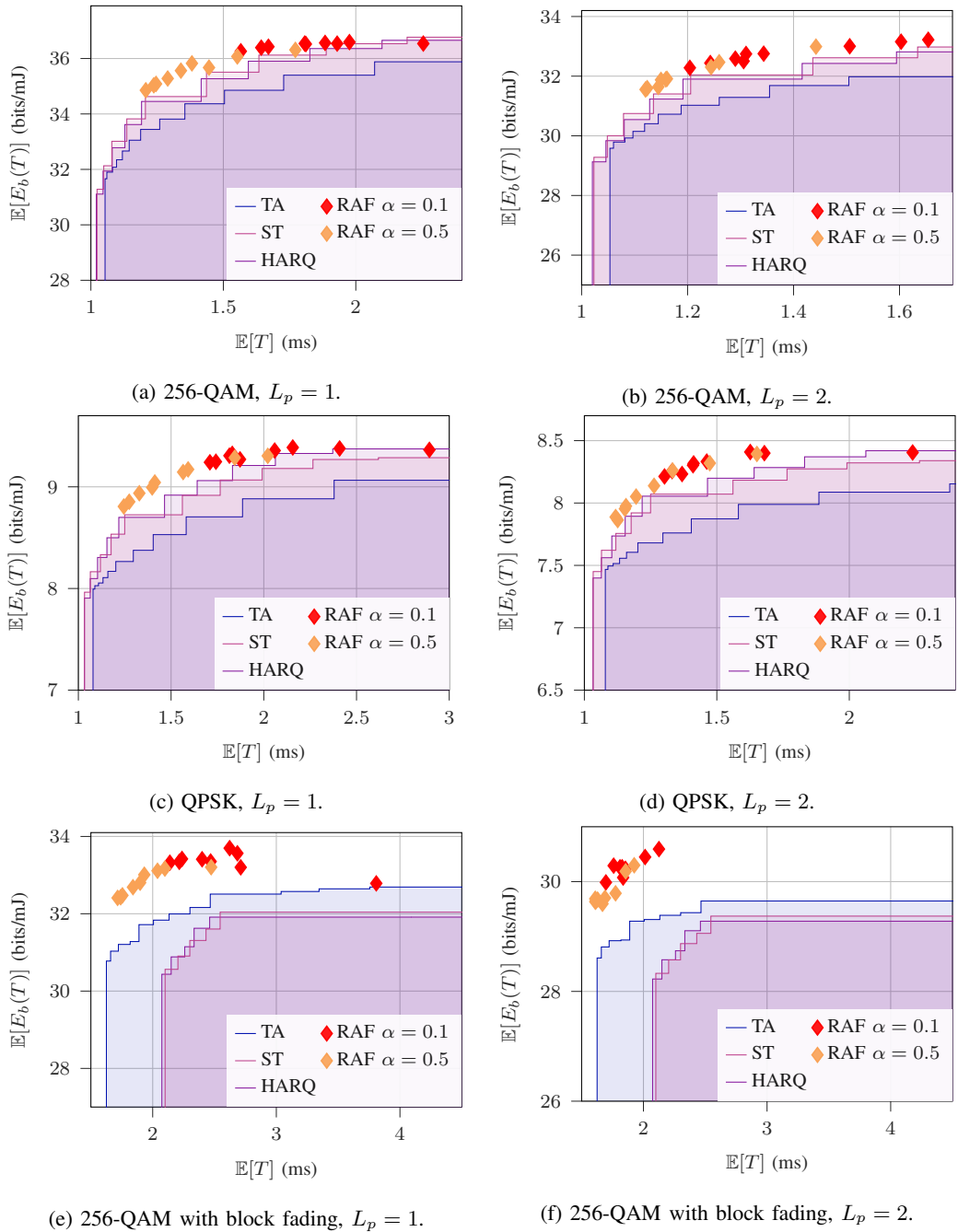

    \centering
        \resizebox{1\linewidth}{!}{%
\begin{tikzpicture}

\pgfplotsset{
    width=\gfwidth,
    height=\gfheight,
    tick align=outside,
    tick pos=left,
    xmajorgrids,
    ymajorgrids,
}

\begin{axis}[
name=plot1,
ylabel style={color=white!15!black,font=\large},
xticklabel style={color=white!15!black,font=\large},
yticklabel style={color=white!15!black,font=\large},
xtick style={color=white!15!black},
ytick style={color=white!15!black},
ylabel={$\mathbb{E}[E_b(T)]$ (bits/mJ)},
xmin=1, xmax=2.39,
ymin=30.1, ymax=37.9,
title= {\large a) 256-QAM, $L_p=1$},
title style={at={(0,0.98)}, anchor=south west}
]
\input{tikzplotlib_figs_icmlcn/256QAM_Pareto_Lp1_data}
\end{axis}

\begin{axis}[
name=plot2,
at={($(plot1.south east) + (26, 0)$)},
ymin=27, ymax=34.2,
yticklabel style={color=white!15!black,font=\large},
ytick style={color=white!15!black},
xmin=1, xmax=1.69,
xticklabel style={color=white!15!black,font=\large},
xtick style={color=white!15!black},
title= {\large b) 256-QAM, $L_p=2$},
title style={at={(0,0.98)}, anchor=south west}
]
\input{tikzplotlib_figs_icmlcn/256QAM_Pareto_Lp2_data}
\end{axis}

\begin{axis}[
name=plot3,
at={($(plot1.south west) + (0, -135)$)},
ymin=7.8, ymax=9.7,
ylabel={$\mathbb{E}[E_b(T)]$ (bits/mJ)},
ylabel style={color=white!15!black,font=\large},
yticklabel style={color=white!15!black,font=\large},
ytick style={color=white!15!black},
xmin=1, xmax=2.99,
xticklabel style={color=white!15!black,font=\large},
xtick style={color=white!15!black},
title= {\large c) QPSK, $L_p=1$},
title style={at={(0,0.98)}, anchor=south west}
]
\input{tikzplotlib_figs_icmlcn/QPSK_Pareto_Lp1_data}
\end{axis}

\begin{axis}[
name=plot4,
at={($(plot3.south east) + (26, 0)$)},
yticklabel style={color=white!15!black,font=\large},
ytick style={color=white!15!black},
xmin=1, xmax=2.39,
ymin=7.2, ymax=8.7,
xticklabel style={color=white!15!black,font=\large},
xtick style={color=white!15!black},
title= {\large d) QPSK, $L_p=2$},
title style={at={(0,0.98)}, anchor=south west}
]
\input{tikzplotlib_figs_icmlcn/QPSK_Pareto_Lp2_data}
\end{axis}

\begin{axis}[
name=plot5,
at={($(plot3.south west) + (0, -135)$)},
ylabel={$\mathbb{E}[E_b(T)]$ (bits/mJ)},
ylabel style={color=white!15!black,font=\large},
yticklabel style={color=white!15!black,font=\large},
ytick style={color=white!15!black},
xmin=1.57, xmax=2.59,
ymin=28.3, ymax=34.1,
xlabel={$\mathbb{E}[T]$ (ms)},
xlabel style={color=white!15!black,font=\large},
xticklabel style={color=white!15!black,font=\large},
xtick style={color=white!15!black},
title= {\large e) 256-QAM with block fading, $L_p=1$.},
title style={at={(0,0.98)}, anchor=south west}
]
\input{tikzplotlib_figs_icmlcn/Fading_test_Pareto_Lp1_data}
\end{axis}

\begin{axis}[
name=plot6,
at={($(plot5.south east) + (26, 0)$)},
yticklabel style={color=white!15!black,font=\large},
ytick style={color=white!15!black},
xmin=1.57, xmax=2.34,
ymin=27.1, ymax=31,
xlabel={$\mathbb{E}[T]$ (ms)},
xlabel style={color=white!15!black,font=\large},
xticklabel style={color=white!15!black,font=\large},
xtick style={color=white!15!black},
title= {\large f) 256-QAM with block fading, $L_p=2$.},
title style={at={(0,0.98)}, anchor=south west}
]
\input{tikzplotlib_figs_icmlcn/Fading_test_Pareto_Lp2_data}
\end{axis}

\matrix [
    draw,
    matrix of nodes,
    anchor=north,
] at ($(plot1.north) + (4.2, 2.35)$) {
    {\large HARQ}   & \ref{plot:HARQ}   & {\large TA}     & \ref{plot:TA} & {\large RAF $\alpha=0.1$}   & \ref{plot:RAFalpha01}  \\
    {\large D-HARQ} & \ref{plot:DHARQ}  & {\large ST}     & \ref{plot:ST} & {\large RAF $\alpha=0.5$}   & \ref{plot:RAFalpha05}  \\
};

\end{tikzpicture}
    }%
    \caption{Pareto frontiers for the average successful bits per mJ versus the average latency in ms for the different baselines and the RAF model when $\alpha=0.1$ and $L_p=1,2$.}\vspace{-0.4cm}
    \label{fig:pareto}
\end{figure}
\def\gfheight {0.9\linewidth}

Finally, we show that \gls{raf} can generalize to other channel by consider the block fading channel $\beta_t \stackrel{iid}{\sim} \mathcal{C}\mathcal{N}(0, 1)$.
The instantaneous \gls{snr} is then $\text{SNR}_t = \frac{|\beta_t|^2}{\sigma^2}$, with a mean \gls{snr} set to $6$ dB. In these experiments, the \gls{raf} models trained on the \gls{awgn} channel are used for testing over the fading channel. The Pareto frontiers are seen in Figs.~\ref{fig:pareto}e and \ref{fig:pareto}f. We can draw two conclusions from these figures: firstly, the static policies cannot adapt to a fading channel since the decisions made by the agents are independent of the instantaneous channel; hence, the \gls{dharq} and \gls{ta} policies are the best baseline policies, while, the \gls{raf} model can generalize to a fading channel maintaining its Pareto dominance. In particular, for a latency of $2$ ms, improvements of $3.9$ and $3.6$ percent over the optimized baselines is observed for the case of $L_p=1$ and $L_p=2$, respectively. An improvement by nearly $4$ percent in energy-efficiency for a battery-powered \gls{iot} device can have significant implications. If for instance the expected lifetime of the battery is $10$ years with the baseline, \gls{raf} increases the lifetime by nearly $5$ months.

\section{Conclusions and Future Work}\label{sec:conc}
In this work, we introduced \gls{raf}, a \gls{dqn}-based approach that enhances \gls{harq} performance for energy-conscious devices in short-packet transmission. By shifting the complexity of adaptive \gls{harq} to the receiver side, the transmitter only needs to decode rich feedback and retransmit the requested symbols. We considered different relative costs for transmitting and receiving feedback, and showed that \gls{raf} with a relatively high feedback cost achieves Pareto dominant policies in the energy efficiency-latency trade-off while having a lower risk of undetected errors. This was observed for both 256-\gls{qam} and \gls{qpsk} over \gls{awgn} and block fading channels.

Future work will explore more complex channels whose statistics can be learned and exploited by the receiver to further optimize the \gls{harq} process.
Moreover, it will be interesting to explore scenarios with imperfect \gls{snr} estimation, other channel codes, and feedback reception errors.

\bibliographystyle{IEEEtran}
\bibliography{bibliography}

\end{document}